\documentclass[%
 reprint,
 amsmath,amssymb,
 aps,
]{revtex4-2}

\usepackage{graphicx}
\usepackage{subfig}
\usepackage{dcolumn}
\usepackage{bm}
\usepackage{caption}
\usepackage{graphicx}
\usepackage{float}
\usepackage{color}
\usepackage{ulem}
\usepackage[modulo]{lineno}

\begin{document}
\preprint{APS/123-QED}

\title{Effective permeability of an immiscible fluid in porous media determined from its geometric state}

\author{Fatimah Alzubaidi}
\author{Peyman Mostaghimi}
\author{Yufu Niu}
\author{Ryan T. Armstrong} \thanks{Corresponding Author: ryan.armstrong@unsw.edu.au}
\affiliation{School of Minerals and Energy Resources Engineering, University of New South Wales, Sydney, Australia}

\author{Gelareh Mohammadi}
\affiliation{School of Computer Science and Engineering, University of New South Wales, Sydney, Australia}%

\author{James E. McClure}
\affiliation{National Security Institute, Virginia Tech, Blacksburg, VA, 24061, USA}%

\author{Steffen Berg}
\affiliation{Shell Global Solutions International B.V., Grasweg 31, 1031HW Amsterdam, The Netherlands}%

\date{\today}

\begin{abstract}
Based on the phenomenological extension of Darcy's law, two-fluid flow is dependent on a relative permeability function of saturation only that is process/path dependent with an underlying dependency on pore structure and wettability. For applications, fuel cells to underground $CO_2$ storage, it is imperative to determine the effective phase permeability relationships where the traditional approach is based on the inverse modelling of time-consuming experiments. The underlying reason is that the fundamental upscaling step from pore to Darcy scale, which links the pore structure of the porous medium to the continuum hydraulic conductivities, is not solved. Herein, we develop an Artificial Neural Network (ANN) that relies on fundamental geometrical relationships to determine the mechanical energy dissipation during creeping immiscible two-fluid flow. The developed ANN is based on a prescribed set of state variables based on physical insights that predicts the effective permeability of 4,500 unseen pore-scale geometrical states with $R^2 = 0.98$.

\end{abstract}

\maketitle

\section{\label{sec:level1}Introduction}
Multiphase flow in porous media is common to various technological applications, including fuel cells \cite{shojaei2022minimal}, supercritical carbon dioxide storage \cite{krevor2012relative}, subsurface hydrogen storage \cite{van2022microfluidics}, contaminant hydrology \cite{falta2005}, recovery of transition fuels \cite{zhang2019synchrotron}, and negative compressibility materials \cite{tortora2021giant}. It is well-known that the constitutive relationship for relative permeability used in multiphase flow models is valid for only a specific process taking a specific saturation path \cite{juanes2006impact,spiteri2008new} and thus, the generality of such models are limited. Processes not captured in that limited experimental parameter space, such as nucleation of a gas phase, e.g., in electrolysis, rather than the co-injection of fluids, result in significantly different relative permeability for the same saturation \cite{gao2021,berg2020determination}. However, there is currently no satisfactory physics-based approach to encode the state space for a generalized relationship even though theoretical works have shown the importance of kinematic analysis beyond saturation only functions \cite{gray2015dynamics}.

We take advantage of a recent discovery that the capillary state at quasi-equilibrium, which is process dependent in a manner similar to relative permeability, can be uniquely defined when a complete set of state variables are considered \cite{mcclure2018geometric}. The approach was based on Hadwiger's theorem that defines the four Minkowski functionals \cite{klain1995} as a complete set of geometrical measures \cite{hadwiger1957vorlesungen}. A piecewise polynomial model (generalized additive model, GAM) can be used to parameterize the state function for quasi-static capillary pressure with the state variables: fluid saturation, surface area, and topology with $0.96 \le R^2 \le 1.0$ \cite{mcclure2018geometric}. The question now is whether this approach can also be extended to effective permeability being a dynamic property based on the viscous stress tensor and velocity field both of which are geometrically constrained by the flowing pathway. While the GAM approach was effective for the fitting of up to three state variables; however, to predict effective permeability, all four Minkowski functionals are presumably required with highly non-linear relationships, which is beyond the possibilities of simple polynomials. As an alternative, artificial neural networks (ANNs) are an attractive candidate for parametrization in higher dimensional space because they are known to be universal approximators \cite{hornik1991approximation}.

The geometrical state of fluids in porous media has been extensively characterised through the utilization of X-ray microcomputed tomography (micro-CT) and microfluidic experiments \cite{blunt2013pore,anbari2018microfluidic}. The earliest microfluidic observations of multiphase flow report different flow regimes from connected pathway flow to ganglion dynamics to drop traffic flow depending on Capillary number ($Ca$) and fluid saturation \cite{avraam1995flow,payatakes1982dynamics}. $Ca$ being the dimensionless
ratio between capillary and viscous forces. For approximately $Ca \le 10^{-5}$, fluids are expected to form connected pathways and provide a linear response between flow rate and pressure drop \cite{zhang2021quantification}. These pathways are geometrically and topologically complex with no analytical solution for the velocity field. This flow regime is the standard for experimental relative permeability measurements in special core analysis \cite{masalmeh2012impact,mcphee2015core}. At $Ca \approx 10^{-5}$, the fluid pathways are commonly reported as being intermittent, resulting in the creation/destruction of interfacial area \cite{reynolds2017dynamic,rucker2015connected}; in addition, a non-linear response between flow rate and pressure drop can be observed \cite{zhang2021quantification,tallakstad2009steady}. The time derivative of the internal energy becomes complex due to additional energy dissipation during geometrical and topological changes and flow rate fluctuations \cite{mcclure2021capillary}. At approximately $Ca \ge 10^{-5}$, the fluids undergo ganglion dynamics whereby disconnected phases flow through the porous domain \cite{payatakes1982dynamics}. 

Presumably, under steady state connected pathway flow, the main contribution to energy dissipation, and thus effective permeability is defined by the viscous stress tensor \cite{raeini2014direct}. The condition of steady state requires that all extensive variables are constant, such as interfacial energy and/or phase volume. Recent works have shown that a detailed balance is maintained under intermittent conditions when time-and-space averaging is considered \cite{mcclure2022relative}. Recent work by \cite{spurin2022red} observed red noise in the pressure signal even at very low $Ca$, suggesting that intermittent filling events occur in a stochastic manner. However, the degree to which the bulk topology of the connected pathways is influenced under low $Ca$ is unclear. Herein, we assume that connected pathways are maintained under creeping flow conditions or at least to a degree at which the impact on relative permeability is negligible. 

We hypothesize that under capillary dominated creeping flow the effective permeability of a phase can be predicted by an ANN when given a complete set of geometric measures, i.e., the Minkowski functionals. Our proposition is that each connected fluid pathway can be characterised by an `equivalent domain' whereby effective permeability, related to total mechanical energy dissipation, is of geometric origin. The approach is inspired by other recent successes with ANN-based models, such as estimation of the stress distribution during finite element analysis \cite{liang2018deep}, force field models to learn the relationship between chemical structure and potential energy \cite{unke2021machine,senior2020improved}, and the identification of Koopman eigenfunctions \cite{lusch2018deep}.

\begin{figure}[H]
\includegraphics[width=1.00\linewidth]{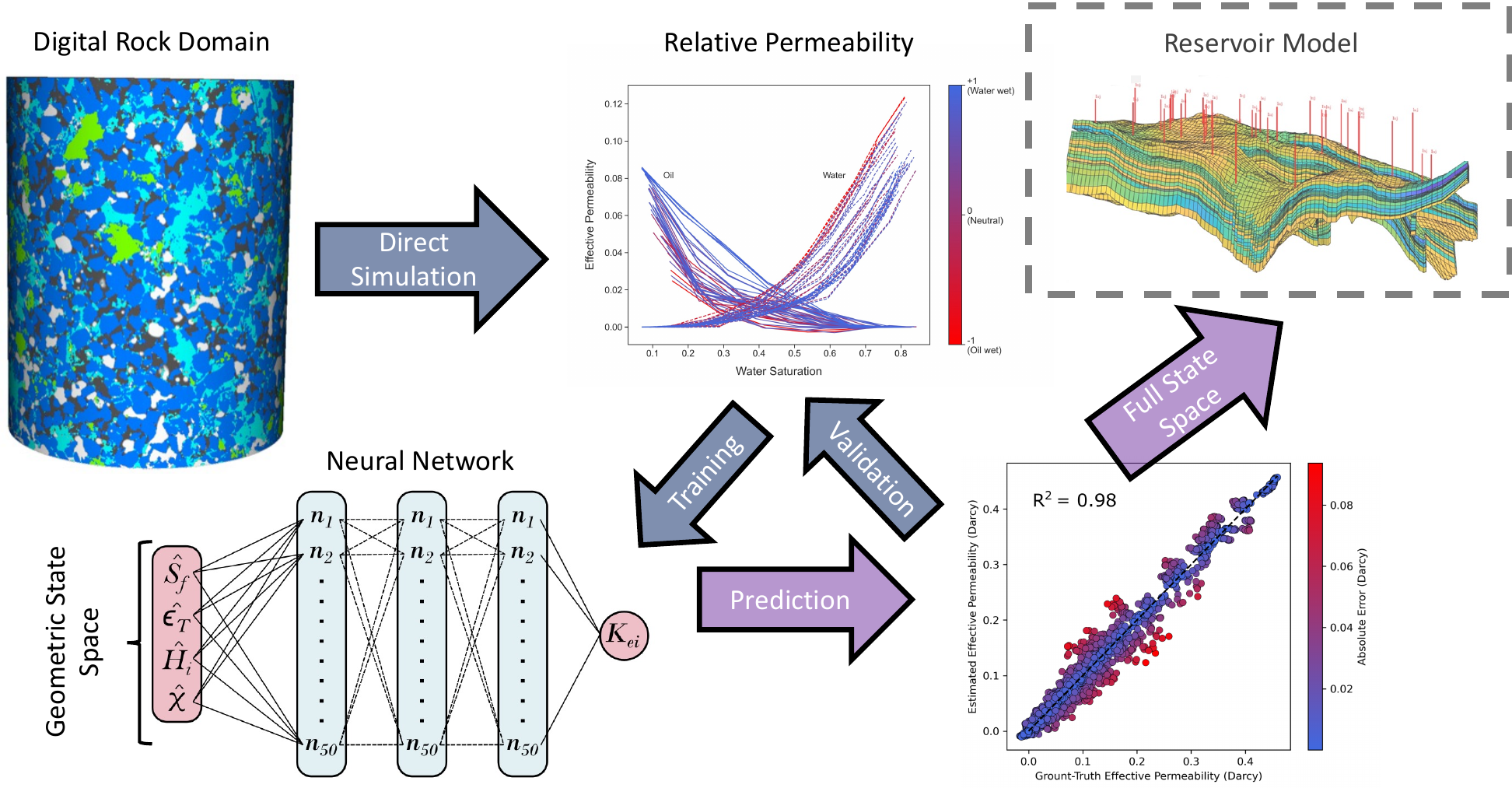}
\caption{Overview of the workflow presented herein and road map for future research on encoding pore-scale information into continuum-scale models by the utilization of artificial neural networks. The grey box around the reservoir model denotes that this step is future work.}
\label{fig:vision}
\end{figure}

The traditional workflow for developing a reservoir model involves the collection of relative permeability data where a single curve represents a single set of parameters defined by Bond number, Capillary number, wettability, and viscosity ratio (or mobility number) and a single history dependence, e.g., imbibition or drainage \cite{mcphee2015core}. The overall workflow proposed herein (Figure \ref{fig:vision}) circumvents these limitations by leveraging the wealth of information provided by pore-scale simulations and recent advances in ANNs to provide a predictive relative permeability for a wide range of operational parameters. The application of an ANN relative permeability model for reservoir simulation remains as future work. Yet the potential for an ANN to predict relative permeability based on the fluid geometrical state alone is the focus of the work presented herein. Our basic approach is to evaluate the impact that the geometrical state of a fluid body has on the relative permeability by using an ANN and pore-scale simulation data. 

\subsection{\label{EDM} Equivalent Domain Model}

To develop our physical insights and provide justification for the state variables, we examine the co-directional flow of two connected pathways through a porous domain. The flow conditions are immiscible, creeping, and capillary dominated. Each connected pathway is decomposed into an `equivalent domain' that contains only the respective phase.

Consider a fluid flow that is subject to force balance
such that Stokes' equations are satisfied locally,
\begin{eqnarray}
  - \nabla p + \nabla \cdot \bm{\tau} &=& 0 \;,
  \label{eq:stokes}\\
   \nabla \cdot \mathbf{u} &=& 0\;,
\end{eqnarray}
where the deviatoric stress tensor is $\bm{\tau}$ and the pressure is $p$. The flow domain $\Omega$ has cross-sectional area $A_c$ and length $L$, and contains multiple fluid regions. 
Let $\Omega_i$ be the region of space occupied by fluid $i$, with an associated
boundary $\Gamma_i$. Each such boundary surface can be decomposed into three regions: the inlet $\Gamma_{ci}(0)$,
the outlet $\Gamma_{ci}(L)$ and the interior boundary $\Gamma_i^{b}$,
\begin{equation}
    \Gamma_i = \Gamma_{ci}(0) \cup \Gamma_{ci}(L)  \cup 
    \Gamma_i^{b}\;.
\end{equation}
In a situation where the internal boundaries do not move, 
the pressure force must balance the contribution of the shear
stresses when considered in the direction normal to the boundary,
denoted by $\mathbf{n}_b$,
\begin{equation}
   \big(p \mathbf{I} - \bm{\tau}\big) \cdot \mathbf{n}_b =  0\;,
   \; \mbox{ (everywhere on $\Gamma_i^b$)}\;.
   \label{eq:5}
\end{equation}
However, this force balance does not hold at the inlet and outlet due to the net flow of fluid $i$ through each surface. The flow is oriented in the same direction as the normal vector $\mathbf{n}$, which will be oriented in a parallel direction for any cross-section. 
Within $\Omega_i$, the flow is Newtonian, 
\begin{equation}
\bm{\tau} = \mu_i  \big( \nabla \mathbf{u}   + (\nabla \mathbf{u} )^T\big)\;,
\quad \mbox{(everywhere on $\Omega_i$)},
\end{equation}
with $\mu_i$ the fluid viscosity. 
If the system is homogeneous, then the volumetric flow rate can be expressed as
\begin{eqnarray}
Q_i =
 s_i  \phi \bar{U}_i A_c = \bar{U}_{si} A_c\;,
    \label{eq:3}
\end{eqnarray}
where $\phi$ is the porosity and $s_i$ is the saturation of fluid $i$. 
Since the fluids are incompressible, the superficial velocity for fluid $i$ will be constant for any 
cross-section $\Gamma_{ci}(z)$, 
\begin{eqnarray}
\bar{U}_{si} = \bar{U}_{si} (z) = \frac{1}{A_c} \int_{\Gamma_{ci}(z)} \mathbf{u} \cdot \mathbf{n} \; {dS} \; \;, 
    \label{eq:4}
\end{eqnarray}
with $z \in [0,L]$.
Equivalently, the volumetric flow rate $Q_i$ is independent of the height along the flow direction $z$. 
\\
We now consider a force balance along the boundary of $\Gamma_i$.
The force exerted on a boundary element includes contributions from the deviatoric stress tensor and the pressure. Considering the whole fluid region, a total force balance requires an equivalence between the 
boundary pressure forces and the internal stresses.
This is stated microscopically by Eq. \ref{eq:stokes}. Integrating this expression over the fluid region $\Omega_i$ and applying the divergence theorem to the pressure term, the pressure difference between the inlet and outlet is linked with the net contribution of viscous forces
within the region $\Omega_i$,
\begin{equation}
 s_i \phi  A_c  \Delta \bar{p}_i 
  = 
\mu_i \int_{\Omega_i}
    \nabla \cdot \big( \nabla \mathbf{u}   + (\nabla \mathbf{u} )^T\big)
     \; dV
     = 0\;,
   \label{eq:6}
\end{equation}
where 
\begin{eqnarray}
\Delta \bar{p}_i =\frac{1}{s_i \phi  A_c} \Bigg(
\int_{\Gamma_{ci}(L)} p \;  dS
 - \int_{\Gamma_{ci}(0)} p \;  dS \Bigg) \;.
 \label{eq:7}
\end{eqnarray}
In general the contributions due to $\bm{\tau} \cdot \mathbf{n}$ will include both interfacial and viscous stresses. In a capillary dominated system, the viscous stresses
are of particular interest since their contribution will account for the quasi-static flow behavior. For a Newtonian fluid, the magnitude for these forces will be determined directly by the pore-scale velocity gradients. 
For porous media the hydraulic radius is commonly used to define a reference length scale as the ratio of cross sectional area to the wetting perimeter, $A_c/L$. The standard definition of hydraulic radius ($r_h$) for an `equivalent' domain is
\begin{equation}
    r_h = \frac{\phi D_p}{6(1-\phi)},
    \label{H_radius}
\end{equation}
where $D_p$ is particle (or grain) diameter and $\phi$ is porosity. 
Assuming a no-slip boundary condition, the flow velocity will go to zero at solid grain boundaries. The hydraulic radius thereby establishes the reference length scale for velocity
gradients. Therefore, we consider the following non-dimensional representation,
\begin{equation}
    \nabla^* \leftarrow {r_h} \nabla\;, \quad
    \mathbf{u}^* \leftarrow \frac{\mathbf{u}}{\bar{U}_{si}} \;.
\end{equation}
For Stokes' flow the structure of the velocity field will be entirely determined from the micro-structure of the porous material.
Conceptually, let us assume that a non-dimensional shape function exists that can account for the net force that is due to the microscopic structure of the viscous stresses,
\begin{eqnarray}
G(\mathbf{shape}) = 
\frac{1}{s_i \phi A_c L} \int_{\Omega{i}
}
\nabla^* \cdot \big( \nabla^* \mathbf{u}^*   + (\nabla^* \mathbf{u}^* )^T\big) 
\; dV \nonumber \\
 \label{eq:8}
\end{eqnarray}
Inserting this into Eq. \ref{eq:6} and
rearranging terms,
\begin{equation}
{\bar{U}_{si}} = \frac{1}{\mu_i}\frac{{r_h}^2 }{ \; G} \frac{\Delta \bar{p}_i } {L} 
  \;.
   \label{eq:9}
\end{equation}
The objective of our artificial intelligence model is therefore to learn the
structure of $G(\mathbf{shape})$, which is expected to depend on invariant
measures of the geometric structure for fluid $i$.

Darcy's law is commonly extended to represent the flow of two immiscible fluids in a porous domain \cite{Muskat1936,Wyckoff1936}. The extended relationship provides a linear relationship for the superficial velocity of phase $i$ as
\begin{equation}
    U_{si} = K_{ei} \frac{\Delta P}{\mu_i L},
    \label{Darcy}
\end{equation}
where $K_{ei}$ is the effective permeability of phase $i$. Based on Equations \ref{eq:9} and \ref{Darcy},
\begin{equation}
    K_{ei} = \frac{\phi^3 D_p^2}{3 G (1-\phi)^2}.
    \label{Ke}
\end{equation}

Effective permeability is therefore based on geometry only. While the geometrical factors are unknown, we know that these parameters are dependent on the morphology of the connected pathway.
\subsection{\label{MFP} Geometrical State Function}
The Minkowski functionals are geometric measures of size based on set theory \cite{serra1983image}. By considering a connected fluid body (X) embedded in Euclidean space ($\Omega$) with boundary surface $\delta X$, integral geometry provides $d+1$ functionals where $d$ is the dimension of $\Omega$. The first functional, $M_0$, is the total volume of $X$. The second functional, $M_1$, is the integral measure of the surface area of $X$. The third functional, $M_2$, is the integral of mean curvature of $\delta X$. The fourth functional, $M_3$, is the integral of Gaussian curvature of $\delta X$.

The Minkowski functionals can be normalized by the total system volume $V$ to provide a set of intensive parameters. The first parameter is simply fluid saturation, $\hat{S_f} = M_o /V \phi$. The second parameter is the specific interfacial area of a fluid body, $\hat{\epsilon_T} = M_1/V$. For the third parameter, an approximation to replace the mean curvature of $\delta X$ with the fluid pressure difference based on the capillary pressure ($P_c$) can be used \cite{mcclure2020modeling}. The approximation is
\begin{equation}
   \hat{H}= \frac{M_2}{V} \approx \frac{P_c}{\gamma} A_{ff}
    + 2 \frac{A_{fs}}{D_s} \;,
    \label{eq:Hi-approx}
\end{equation}
where $A_{ff}$ is the fluid/fluid interfacial area and $A_{fs}$ is the fluid/solid interfacial area. Conveniently $P_c$ is already defined in multiphase flow models. The interfacial tension $\gamma$ is usually a known parameter, and the grain Sauter diameter $D_s$ is directly measurable. The fourth parameter provides the degree of connectivity per volume, $\hat{\chi} = M_3/V$. 

Hadwiger's theorem states that any linear additive property ($F$) of an object can be represented by a linear combination of the Minkowski functionals. \cite{hadwiger1957vorlesungen}
\begin{equation}
F(X) = \sum_{i=0}^{3} c_i M_i(X),
\label{Hadwiger}
\end{equation}
where $c_i$ is a coefficient that is independent of fluid body (X). Motion-invariant means that $F(X)$ does not depend on the orientation of $X$. 

Therefore, Equation \ref{Ke} is reformulated as 
\begin{equation}
K_{ei} = f(\hat{S_f}, \hat{H_i}, \hat{\epsilon_T}, \hat{\chi}).
\label{RE}
\end{equation}
The independent parameters are commonly found in porous media multiphase flow research \cite{armstrong2016beyond,liu2017pore,scholz2012permeability}. The task herein is to define Equation \ref{RE}. 

\section{\label{sec:lMM}Materials and Methods}

\subsection{\label{MFP} Multiphase Flow Simulations}
Simulations were conducted using a parallel implementation of the colour-gradient Lattice Boltzmann method (LBPM) \cite{mcclure2021lbpm}. The code has been documented, validated, and tested in numerous publications \cite{McClure_Berrill_etal_16,mcclure2014novel,armstrong2016beyond,liu2018influence,mcclure2018geometric}. To obtain a diverse set of geometrical states under creeping flow conditions, we simulated a wide range of wetting conditions \cite{armstrong2021multiscale}. In total, the simulations took 12 hours of walltime using 48 NVIDIA V100 GPUs on the Summit Supercomputer.

We used a segmented micro-CT image of a North Sea sandstone where the mineral phases were identified on a per voxel basis. The details of the image processing workflow and mineralogy are provided in the Supporting Materials. For simplification, we grouped the mineral voxels into three groups - quartz (74\%), clay (19\%) and carbonate (6.3\%). For initializing numerical simulations, the mineral groups were defined by various contact angles that represent a wide range of different fluid pairs. The wettability of these systems was defined as a summation of the cosine of the contact angles for each mineral/fluid/fluid combination, determined as 
\begin{equation}
\begin{split}
W = 
(\gamma_{qn}-\gamma_{qw})/\gamma_{wn} \phi_{q} +
(\gamma_{kn}-\gamma_{kw})/\gamma_{wn} \phi_{k} + \\
(\gamma_{cs}-\gamma_{cs})/\gamma_{wn} \phi_{c}
\label{eq:wetindex}
\end{split}
\end{equation}

where $\phi_{q}$, $\phi_{k}$ and $\phi_{c}$ are the solid voxel fractions of quartz, clay and carbonate, respectively. This metric provides a scale from strongly water-wet ($W = 1.0$) to strongly oil-wet ($W = -1.0$). In addition, the squared variance, $\sigma^2$, of the contact angle values defines the spatial variability. In total 36 different wetting conditions were simulated. The wetting distribution was considered in three ways. (1) The mineral type is the only reason for the wetting affinity to vary. (2) Fluid history is the only reason for the wetting affinity; based on an initial morphological drainage of the image to 20 percent saturation. (3) Wetting affinity varies based on both fluid history and mineral type.

Further details on the image processing, wetting distributions, and contact angles (affinity values) used are provided in the Supporting Materials.

\begin{figure}
\includegraphics[width=0.9\linewidth]{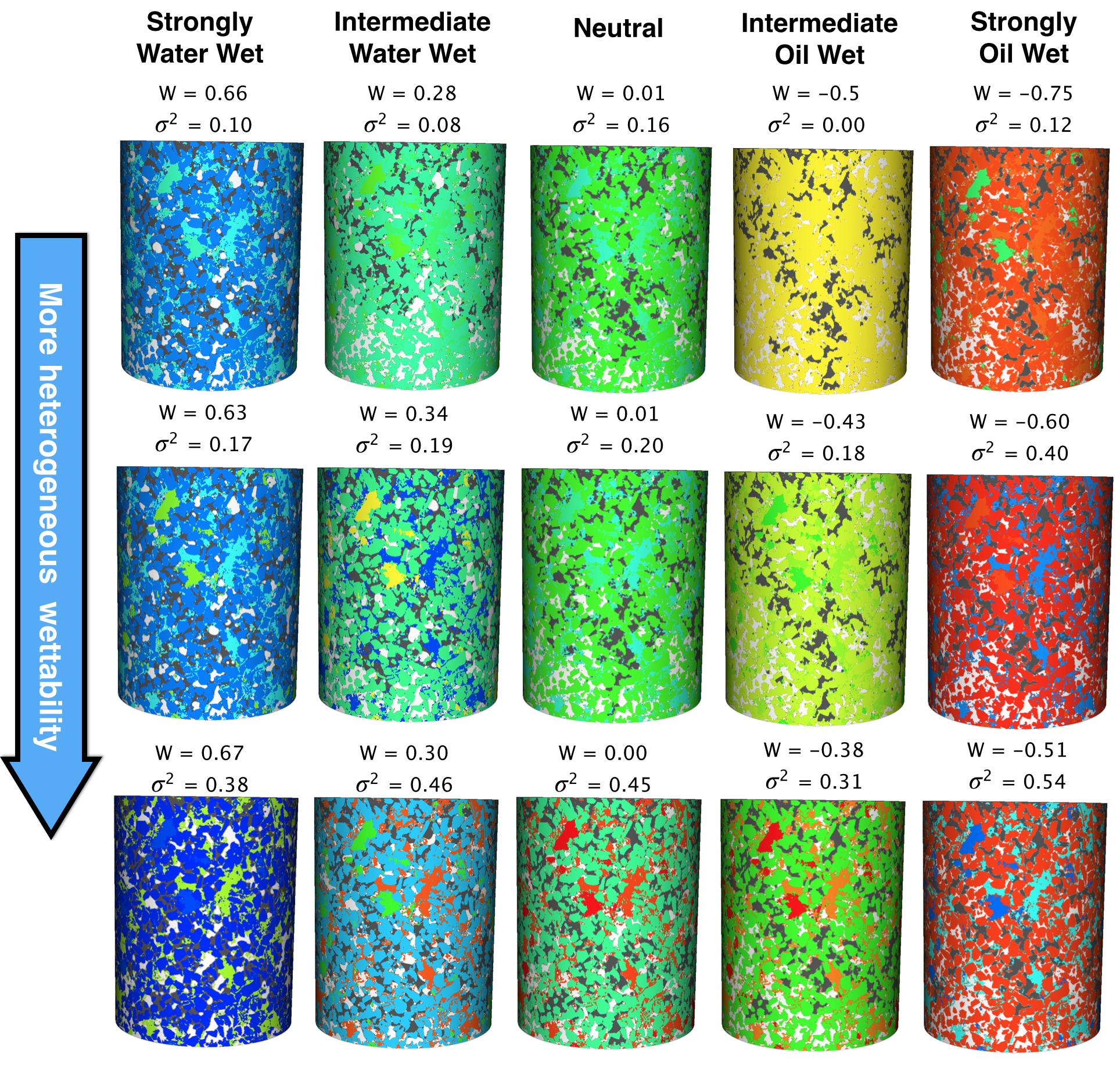}
\caption{Distribution of oil (white) and water (black) obtained from simulations of water-flooding under different wettability conditions. The solid material is colored according to the local wettability ranging from water wet (blue) to oil wet (red).}
\label{fig:wettability-distribution}
\end{figure}

\subsection{\label{ANN} Artificial Neural Network}
A simple ANN was tasked to find Equation \ref{RE} within the defined state space. The ANN had three hidden layers, each with 50 neurons and was trained for a total of 250 epochs using the stochastic gradient decent optimizer with an initial learning rate of 0.001, which was reduced by 10\% after 10 and 150 epochs. Further details on the network optimization, training, and testing are provided in Supporting Materials.

\section{\label{sec:RD} Results and Discussion}
In Fig. \ref{fig:wettability-distribution} the steady state fluid distributions obtained for various wetting conditions are presented. The simulation conditions ranged from strongly water wet to strongly oil wet, which was quantified by the average wettability of the system ($W$) and mineral heterogeneity based on the squared variances ($\sigma^2$) of the wettabilities assigned to these minerals. The solid material is colored based on the local range of wettability with water-wet solid being blue and oil-wet solid being red. The oil phase is white and water is black. It was observed that the distribution of fluids was dependent on both $W$ and $\sigma^2$. In particular, for similar $W$ (pick a column in Fig. \ref{fig:wettability-distribution}) the distribution of fluids differs based on how wettability was assigned to the spatial distribution of the minerals. This result is well aligned with other recent studies; concluding that the spatial distribution of wetting is an important parameter for the simulation of multiphase flow in porous media \cite{foroughi2020pore,foroughi2021pore,garfi2022determination}. This also suggests that pore space geometry alone is insufficient for the prediction of relative permeability. 

\begin{figure}[H]
\includegraphics[width=1.00\linewidth]{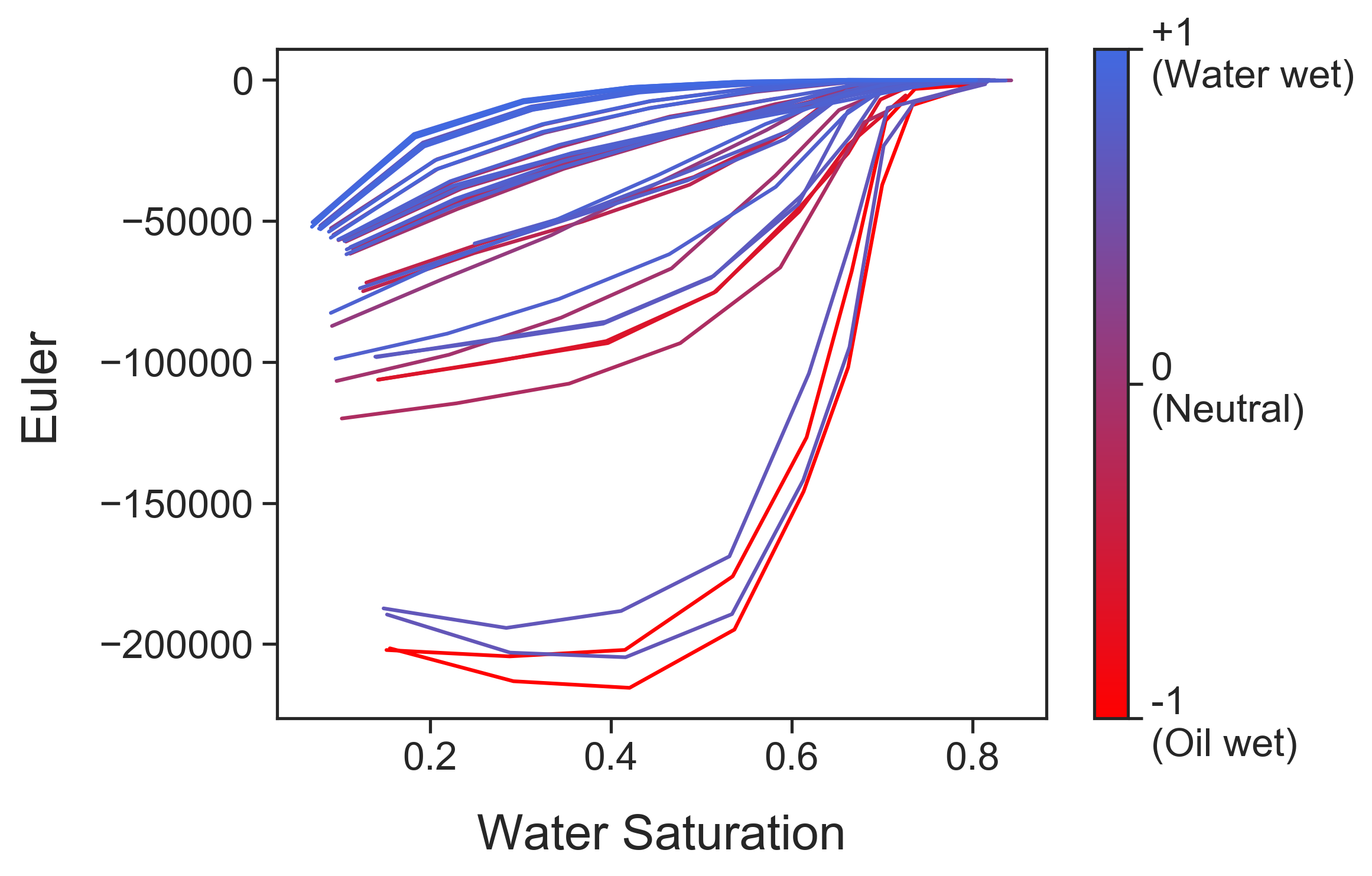}
\caption{LBPM Simulations provided 36 different Euler characteristic trends for each wetting case.}
\label{fig:euler}
\end{figure}

\begin{figure}[H]
\includegraphics[width=1.00\linewidth]{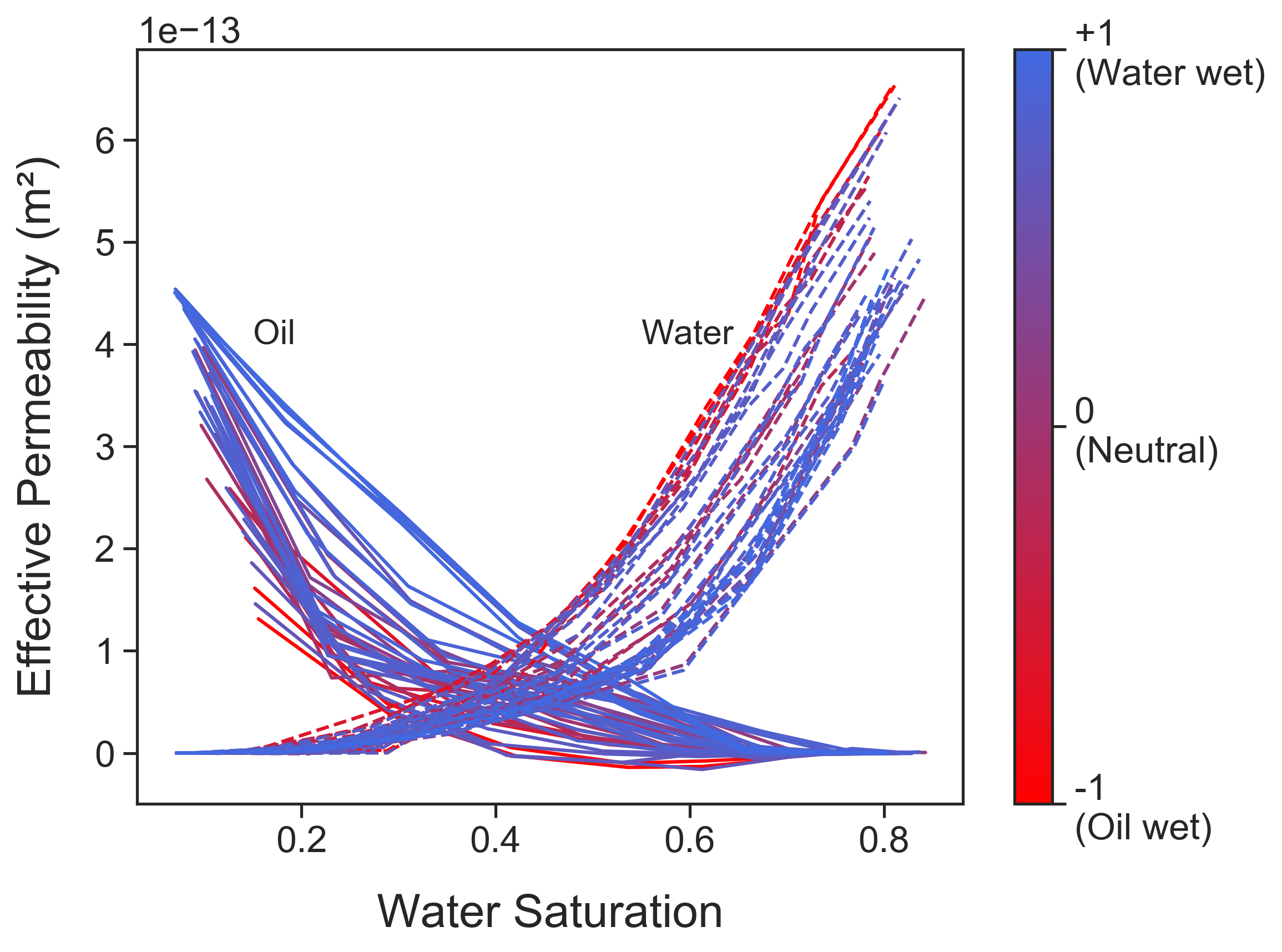}
\caption{LBPM Simulations provided 36 unique relative permeability curves for each wetting case.}
\label{fig:rel_perm}
\end{figure}

The LBPM simulations provided 25,906 unique morphological states. Euler characteristic for the (non-wetting phase) NWP followed the expected trend of increasing connectivity for more oil-wet conditions (Figure \ref{fig:euler}). The NWP stops percolating at high water saturation near Euler equal to one resulting in an effective permeability of approximately zero at the corresponding saturation. The constitutive relationship for relative permeability was well aligned with the expected behaviour based on wettability (Figure \ref{fig:rel_perm}). The oil phase effective permeability increased under more water-wet conditions while the water phase relative permeability decreased resulting in a net effect of shifting the crossing point of the two curves to the right. As evident, no single constitutive relative permeability relationship would be sufficient to represent this data. 

\begin{figure}
\includegraphics[width=1.00\linewidth]{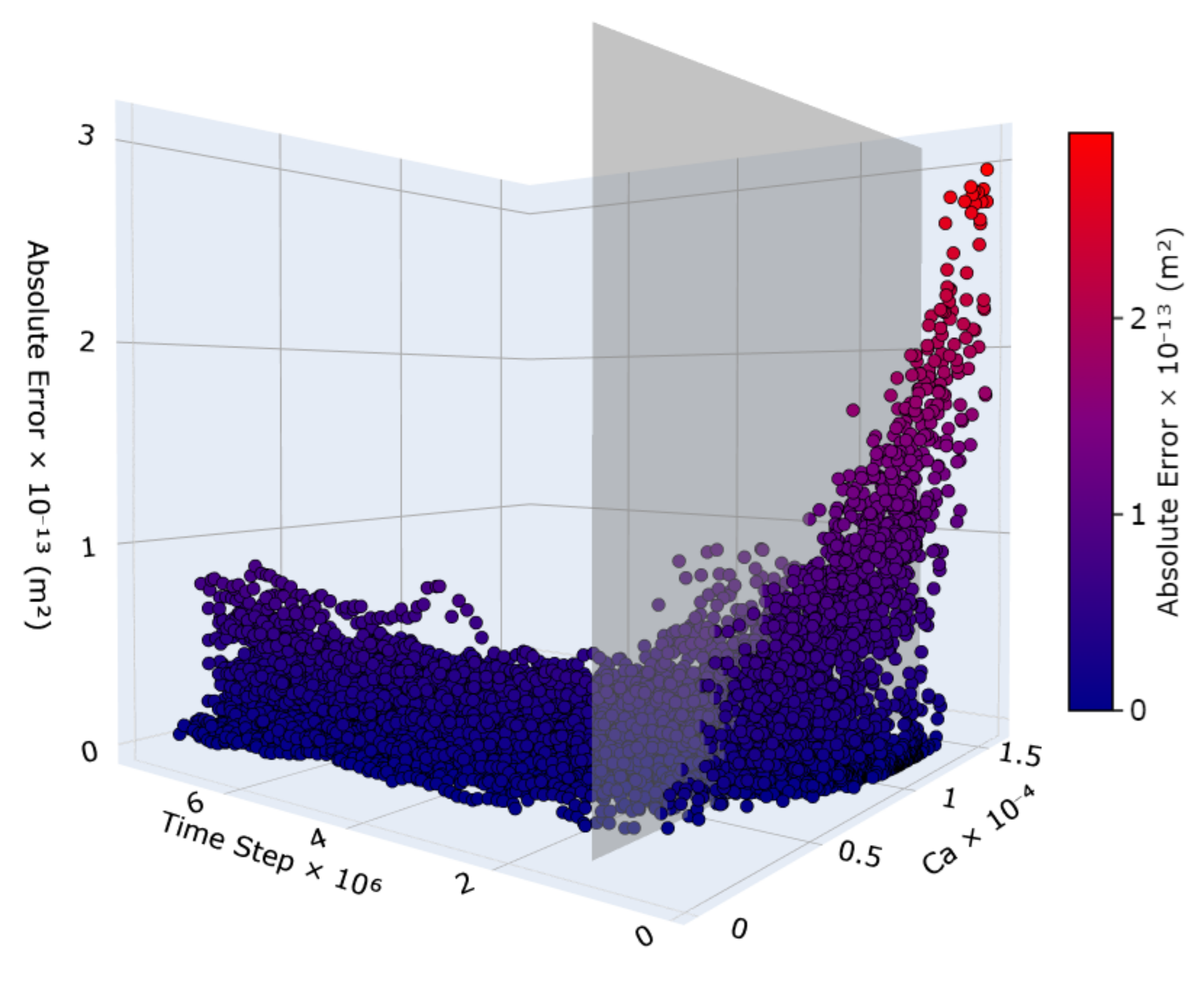}
\caption{Representative points from the simulation obtained after $1\times10^6$ time steps (the grey plain). Data after this time step was used to build the ANN model presented in the main paper.}
\label{fig:data_selection}
\end{figure}

Based on the creeping flow assumption underpinning Equation 15, the effective permeability of a phase can be expressed as a function of geometry only. Our premise is that an ANN should be ably to find the mapping between Equation 15 and Equation 18, as suggested by Hadwiger's theorem. The LBPM simulations provided a total of 25,906 data points. We found that representative points were obtained after the simulation convergence, which can be indicated by a convergence of $Ca$. Therefore, by visualizing $Ca$ over simulation time in addition to the associated error for each point (see Figure \ref{fig:data_selection}), we have included points after $1\times10^{6}$ time steps. The excluded points represent the earliest set recorded during the simulation at high $Ca$, as data was recorded every $1\times10^{6}$ steps to up to $78\times10^{5}$. The remaining data at low $Ca$, which include 22,556 points, was used to develop and evaluate the ANN.

In Figure \ref{fig:data_selection} the ANN absolute errors associate with all 25,906 geometric states are provided along with Capillary number, $Ca$, and the simulation time step. Low $Ca$ is associated with creeping flow conditions as required for Equation 15. For an ANN to find the mapping between Equation 15 and 18, and thus predict effective permeability, the effective permeability must be based on the geometry of the connected pathway only, this situation is commonly refereed to as connected pathway flow \cite{avraam1995flow}. As evident in Figure \ref{fig:data_selection}, earlier simulation time steps, associated with higher $Ca$, resulted in higher absolute error suggesting that the under pinning assumptions no longer hold. 

For higher $Ca$ flows, the time-dependent aspect of the phenomena cannot be captured by geometrical terms expressed in units of length only; the fluid pathways are intermittent with geometrical and topological changes \cite{reynolds2017dynamic}, and thus energy dissipation depends on the geometric state and the fluctuation timescale \cite{mcclure2021capillary}, resulting in high ANN absolute error. While a unit of time can be introduced by $Ca$, the addition of $Ca$ as an input parameter into the ANN did not improve the predictive capability of the ANN - see Supporting Materials. In addition, we do not present any theoretical reason as to why $Ca$ should be included. 

\begin{figure*}
\includegraphics[width=1.0\linewidth]{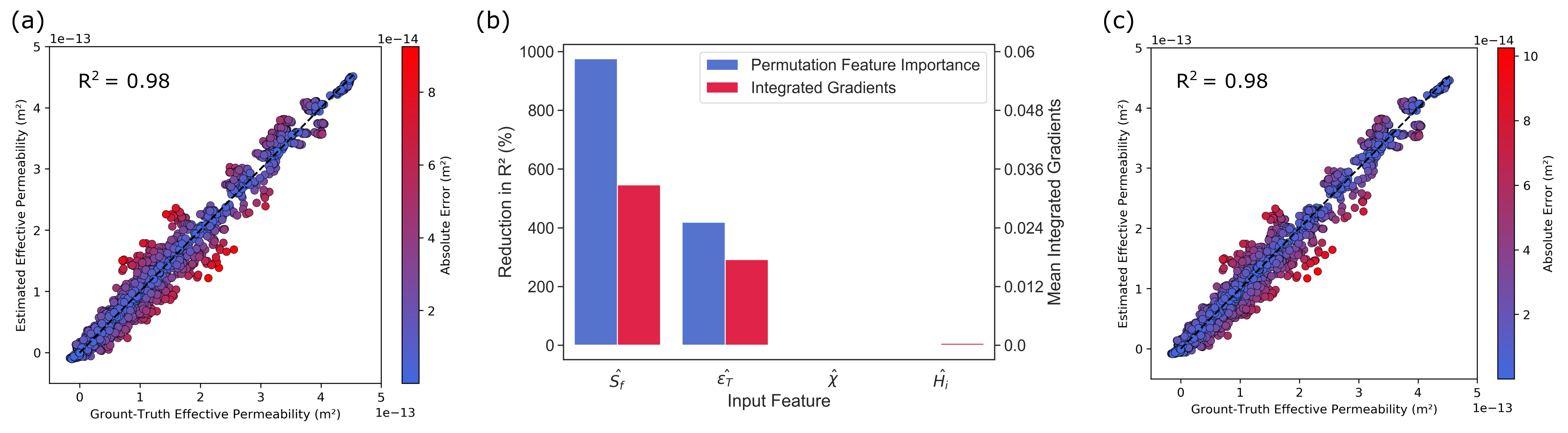}
\caption{(a) ANN predictability on the test set with all 4 Minkowski functionals as input. (b) The importance of each input parameter for the result presented in (a). (c) A reduced order model, $K_{ei} = f(\hat{S_f}, \hat{\epsilon_T})$, with the same predictability as the full model presented in (a).}
\label{fig:results}
\end{figure*}

To explore capillary dominated connected pathway flow, we included only simulation data after $1\times10^{6}$ time steps, resulting in 22,556 geometric states. The results for 4,500 unseen geometric states from the testing set are provided in Figure \ref{fig:results}(a). The network achieved an average absolute error of less than $9.87 \times 10^{-15} m^2$. Approximately 68\% of the test points have an absolute error below the average while only 3\% have an error of more than $3.95 \times 10^{-14} m^2$, up to a maximum error of approximately $9.87 \times 10^{-14} m^2$. The results indicated a valid relationship for the geometric state of the connected pathway flow and the effective permeability. 

Feature importance was evaluated using the permutation feature importance and integrated gradients methods. The importance of each feature is presented in Figure \ref{fig:results}(b) were the importance of the last two features were negligible. The permutation feature importance method \cite{Breiman2001} evaluates the change in the predictability of an ANN resulting from randomly rearranging the order of the measured feature and/or providing only the mean value for a given feature. As the accuracy of the model was measured using $R^{2}$, the importance was evaluated as the percentage reduction in the baseline $R^{2}$. The integrated gradients (IG) method \cite{Sundararajan2017} estimates feature importance for a trained model by accumulating the gradients along a path between the input and a baseline input. Only the absolute value of the IG has been shown in Figure \ref{fig:results}(b) as a measure of importance. Based on these methods, the most important features were $(\hat{S_f}, \hat{H_i})$ while $(\hat{\epsilon_T}, \hat{\chi})$ had little to no influence on the the ANN predictability. Therefore, a reduced order model based on $K_{ei} = f(\hat{S_f}, \hat{H_i})$ as presented in Figure \ref{fig:results}(c) could be generated with the same predictability as the original model.

\section{\label{sec:level1} Conclusions}
In principle, our previous works on capillary pressure \cite{mcclure2018geometric,mcclure2020modeling} were easier to deal with since capillary pressure is a quasi-static property where one can consider static snap-shots with no flow and fluid-fluid interfaces are near equilibrium conditions, and thus capillary pressure is related to the mean curvature of the fluid interfaces \cite{armstrong2012linking}, which is one of the Minkowski functionals. Therefore, capillary pressure, is fundamentally a geometrical statement. With effective permeability (or relative permeability) we need to extend our concepts to flowing conditions. We therefore concentrated on the capillary-dominated regime, which is typically relevant for reservoir engineering and also covered by special core analysis (which does not extent above $Ca = 10^{-5}$, see e.g. \cite{masalmeh2012impact}). Our overall approach was to relate the mechanical energy dissipation (net force) to the geometric boundary of the connected pathway, or an ‘equivalent’ boundary with a given shape factor. 

The determination of effective permeability from geometrical measures suggests that any two systems with the same effective permeability have the same total energy dissipation (net force) under stationary capillary dominated conditions. While the Minkowski functionals are a complete set they are not a unique set \cite{hadwiger1957vorlesungen}. Therefore, our results suggest that the simulations provided structures for all wetting states where the flow field retained a degree of similarity near a critical percolation \cite{scholz2012permeability}. Therefore, within a given class of conditions effective permeability models can be linked to energy dissipation (or net force), which is additive, motion invariant, and smooth, as required by Hadwiger \cite{hadwiger1957vorlesungen}. We  should also emphasize anisotropic extensions for future work, since there are many more invariants when you consider direction-dependence \cite{schroder2011minkowski}. Since relative permeability is a tensor in practice, the addition of directional dependencies should be an area of future work. 

Our previous works have shown that the capillary pressure curve can be predicted from the Minkowski functionals with two geometric degrees of freedom when a non-dimensional form is constructed \cite{mcclure2018geometric,mcclure2020modeling}. For the van Genuchten relative permeability model the behaviour is parameterized based on the capillary pressure curve \cite{van1980closed}, which supports the reduced order model found herein. Other works have suggested the importance of phase connectivity \cite{purswani2021predictive,liu2017pore}, however, these works were under strongly water-wet conditions while the current study is intermediate/mixed wet \cite{foroughi2020pore} with a high degree of connectivity and minimal surfaces \cite{lin2019minimal}. Spatially varied wetting, $\sigma^2$, essentially provided a degenerate state whereby a limited set of geometrical measures were required, namely saturation and surface area only, which further supports the need for surface wettability characterisation \cite{garfi2022determination} in pore-scale models as necessary input (initial conditions include the spatial wettability distribution, at least in a statistical manner). 

A consequences of the phenomenological two-phase Darcy formulation is that effective permeability needs to be experimentally measured, which is often determined from inverse modelling that is ill-posed with non-uniqueness, correlation between matching parameters, and non-Gaussian errors \cite{berg2021non}. The associated uncertainty by inverse modeling is greater than that determined by the presented ANN \cite{berg2021non,berg2021sensitivity}. In addition, the determination of end-point relative permeability is nebulous. Experimental methods for determining end-point relative permeability are based on 'Best Practices' \cite{mcphee2015core}, and thus trends between end-point relative permeability and wetting index are often conflicting \cite{christensen2018residual,jadhunandan1995effect}. End-point relative permeability prediction with an ANN is based on simulation data with difficulties similar to experimental data were a significant number of pore volumes (or time steps) at an increased flow rate are required to achieve residual saturation \cite{mcclure2021lbpm}. This is particularly the case for intermediate wet conditions under which thin wetting films remain connected with low hydraulic conductivity \cite{armstrong2021multiscale}. 

The proposed ANN approach requires a significant amount of pore-scale data for training, which is an upfront computational cost. Morphological data, however, is becoming readily available with recent developments in pore-scale modelling and experiments \cite{blunt2013pore} and data repositories, such as the 'Digital Rocks Portal'. The presented ANN was able to predict the effective permeability for 36 unique wetting states. With traditional 
constitutive models, a steady state core flood would need to be conducted for each wetting state with each experiment taking approximately one week to conduct. Such an enormous parameter space (as that explored by the ANN) is not feasible with traditional experimental workflows and subsequent constitutive models.

Our ANN was able to predict effective permeability with $R^2=0.98$ for a wide range of different wetting states within the same rock. Corey and/or van Genuchten relative permeability models would require different parameterisation for each wetting case \cite{brooks1964hydraulic,van1980closed}. The extrapolative power of an ANN is not without bounds. Analogous to how a Corey model is limited to a defined set of control parameters(experimental conditions), the ANN is limited to the features of the training data \cite{da2021deep}. Application of the ANN to a wider set of control parameters would need to be explored in future work. Currently, the ANN provides excellent results for a single rock type under a wide range of wettabilities, which is a considerable improvement over traditional constitutive relationships since traditional models are limited to a single wetting condition. 

Our ANN provides a physics-based approach to predict effective permeability by only geometrical descriptors providing new possibilities for experimental design and multiphase modelling. Typical multiphase models already include two of the parameters ($\hat{S_f}, \hat{H_i}$) proposed in Equation 18 while recent models include $\hat{\epsilon_T}$ \cite{hassanizadeh1993thermodynamic,gray2015dynamics} and the inclusion of $\hat{\chi}$ is not far off \cite{mcclure2018geometric}. ANNs as constitutive models are attractive because they will significantly benefit from hardware acceleration over the near future. An ANN could realistically be embedded into a reservoir simulator and used to model significantly more complex physics without having it dramatically slow things down. Similar approaches are already being used for collision operators in lattice Boltzmann models and molecular dynamics simulations \cite{bedrunka2021lettuce}.

\begin{acknowledgments}
R. T. A. acknowledges Australian Research Council Future Fellowship (FT210100165) and Discovery (DP210102689). The authors thank Mehdi Shabaninejah for kindly sharing the micro-CT and QEMSCAN data.
\end{acknowledgments}

\bibliography{apssamp}

\end{document}